\documentclass[aps,prb,twocolumn,superscriptaddress,floatfix,nofootinbib]{revtex4-2}

\usepackage{amsmath,amssymb,bm,amsthm}

\usepackage{graphicx}
\usepackage[caption=false]{subfig}
\usepackage{xcolor}
\usepackage{appendix}
\usepackage{hyperref}
\hypersetup{
    colorlinks=true,
    linkcolor = [rgb]{0.70,0.13,0.13}, 
    citecolor = [rgb]{0.13,0.55,0.13},
    urlcolor = [rgb]{0.25, 0.41, 0.88} , 
    filecolor=cyan,      
    pdfcreator = {\LaTeX\ and \flqq hyperref\frqq},
}
\usepackage{caption}
\usepackage{subcaption}
\usepackage{cleveref}
\usepackage[vcentermath]{youngtab}
\usepackage{upgreek}
\usepackage{bm,dsfont}
\usepackage{multirow}
\usepackage{braket}
\usepackage{enumitem}
\usepackage[percent]{overpic}
\usepackage{bbm}
\usepackage{bm,dsfont}
\usepackage{comment}
\usepackage{endnotes}
\graphicspath{ {./Figs/} } 
\newcommand{\cm}{c_-}
\newcommand{\JJ}{J(A,B,C)}
\newcommand{\kv}{\bm{k}}
\newcommand{\dv}{\bm{d}}
\newcommand{\sv}{\bm{\sigma}}
\newcommand{\ch}{\mathcal{C}}
\newcommand{\Tr}{\mathrm{Tr}}
\newcommand{\Ql}{\mathcal{Q}_{<}}
\newcommand{\Qg}{\mathcal{Q}_{>}}

\newcommand{\dm}{\delta m}

\begin{document}
\title{Parity Anomaly as Modular Commutator with Massless Dirac Fermion}

\author{Meng Zeng}
\affiliation{Max Planck Institute for the Physics of Complex Systems, Nöthnitzer Str. 38, 01187 Dresden, Germany}

\date{\today}

\begin{abstract}
The modular commutator $\JJ = i\langle[K_{AB},K_{BC}]\rangle$ extracts the chiral central charge $\cm$ from a single bulk wavefunction of a \emph{gapped} 2d state, where $3J/\pi=\cm$. Inspired by the recent developments in the field of gapless symmetry-protected topological phases,  we ask: what does the modular commutator measure, if it is well-defined at all, when the 2d bulk becomes \textit{gapless}? Several interesting new insights can already be obtained using the simple Haldane honeycomb model. For the critical point hosting an isolated Dirac node we find that $J$ remains sharp: it converges to a \textit{half-quantized} value, with corrections that decay as a power law in the subsystem size rather than exponentially, mirroring the power-law correlations in gapless systems. We prove the half-quantization using an emergent reflection symmetry of the massless Dirac cone, and show that the half-quantized contribution comes from the other gapped cone (the massive partner of the massless one). This massive partner can be interpreted as the physical incarnation of the Pauli-Villars regulator, which is the origin of the parity-breaking level-$\frac{1}{2}$ Chern-Simons term (with half-quantized Hall conductance) and the parity anomaly. When protected chiral edge modes coexist with a bulk Dirac node we obtain $3J/\pi = \cm+\tfrac12$. The half-quantization is also shown to be robust against tripartition deformation, tuning Dirac velocity and Dirac cone anisotropy. We further investigate other types of gaplessness---quadratic nodes (in contrast to linear Dirac) and the case with Fermi surface---and show that the robust half-quantization of $J$ is lost in such non-Dirac cases. These results generalize the modular commutator beyond gapped phases, and at the same time provide an information-theoretic measurement of the parity anomaly. 
\end{abstract}

\maketitle

\section{Introduction}
Phases of matter beyond the Landau paradigm are difficult to characterize by local order parameters. 
Among other available characterizations, information-theoretic approaches have been powerful tools in studying such exotic phases.
The topological entanglement entropy---a constant, subleading correction to the entanglement area law that measures the total quantum dimension---is a canonical example~\cite{LevinWen2006,KitaevPreskill2006}.
The entanglement spectrum diagnosis is another classic example~\cite{LiHaldane2008}.
Such developments and the related endeavors to fully classify gapped topological phases in arbitrary dimensions have motivated the recent proposal of the so-called ``entanglement bootstrap'' program~\cite{ShiKatoKim2020,shi2021DW,shi2021DW-TEE}, which aims to extract all the topological information, including the total quantum dimension and the anyon fusion rules etc., solely from the entanglement properties of the ground state wavefunction, with a minimal entanglement area law assumption. 

One consequence of this program is the discovery of a clean formula for one piece of important topological data, the chiral central charge (CCC) $c_-$, using \textit{only} the ground state wavefunction~\cite{KSKA2022}. 
The CCC fixes the quantized thermal Hall conductance, counts the net chirality of the boundary modes, and obstructs any symmetric gapped boundary due to the gravitational anomaly. This formula is based on the modular commutator (MC), defined for a tripartition of a subregion~\cite{KSKA2022,KSKA2022long}, which involves the commutator of the modular/entanglement Hamiltonians of overlapping subregions. Therefore, once the ground state is known, the MC, denoted by $J$, can in principle be calculated, and it has been proven that for generic 2d gapped area-law states $J=\frac{\pi}{3}c_-$. This relation can be interpreted in terms of modular flow (the dynamics generated by the modular Hamiltonian~\cite{Zou2022,Fan2022}), and has also been verified in interacting spin models~\cite{maity2025identifying}. It is worth mentioning that there can be outliers that violate this relation when the modular Hamiltonians become nonlocal~\cite{levin2024}.

All the previous discussions related to the MC presuppose a gapped bulk with area law entanglement, which is believed to hold for generic gapped states, even though it is not yet rigorously proven for dimensions higher than one~\cite{hastings2007area}. 
The original argument relating the MC and the $c_-$ rests crucially on the area law (and derived properties thereof), which can be potentially modified in the presence of gapless degrees of freedom. It is therefore not obvious what happens once the gap closes. 
This is a timely question to ask, in light of the recent developments in the field of gapless symmetry-protected topological
(SPT) phases, where robust symmetry-protected edge modes appear in the presence of a gapless bulk~\cite{yu2026review} and where many conventional topological invariants become ill-defined. In such cases, is the MC still well-defined? If so, what does it measure? Is it still able to capture the signature of the chiral edge modes?

In this work we take a first step to go beyond gapped phases by addressing these questions in a simple controlled setting: the free-fermion Haldane honeycomb model tuned to its transition, where the
relevant reduced density matrices follow directly from fermion correlation functions that are straightforwardly calculable even at large
system sizes~\cite{Peschel2003,PeschelEisler2009}. Our main findings are the following:
\begin{enumerate}
\item[(i)] \emph{The MC is half-quantized for Dirac nodes.} When a topologically trivial bulk consists of one massless Dirac cone, we find $3J/\pi = \tfrac12$, with power-law finite-size corrections. We give a symmetry-based argument for the half-integer quantization and interpret it as the entanglement counterpart of the parity anomaly. In this case, the entanglement area law still holds, but we will comment that it is a different type of area law compared to the original entanglement bootstrap program in the gapped case. This difference is reflected in the violation of Markov property of chain tripartitions~\cite{ShiKatoKim2020,KSKA2022,KSKA2022long}. The half-quantization is robust against tripartition deformation, tuning Dirac velocity and anisotropy of the Dirac cone.
\item[(ii)] \emph{Bulk and edge contributions do not directly separate.} When a protected chiral
  edge modes coexist with a bulk Dirac node as in the gapless SPT case, $3J/\pi = c_-+ \tfrac12$, with $c_-$ indicating the net chirality of the edge. That means, $J$ responds to the total chirality of the system.
\item[(iii)] \emph{Dirac node is special among the different types of gaplessness}. Systems with other types of gapless bulk, e.g. quadratic node or Fermi surface, no longer have robust MC. The MC of the quadratic node depends on the curvature of the node, even though area law still holds and the MC is invariant under deformation of the tripartition. In contrast, the Fermi surface already violates the area law, and its MC becomes strongly tripartition-dependent and is no longer well-defined.

\end{enumerate}

\section{Setup and model}
\label{sec:setup}

\begin{figure}
    \centering
    \includegraphics[width=\linewidth]{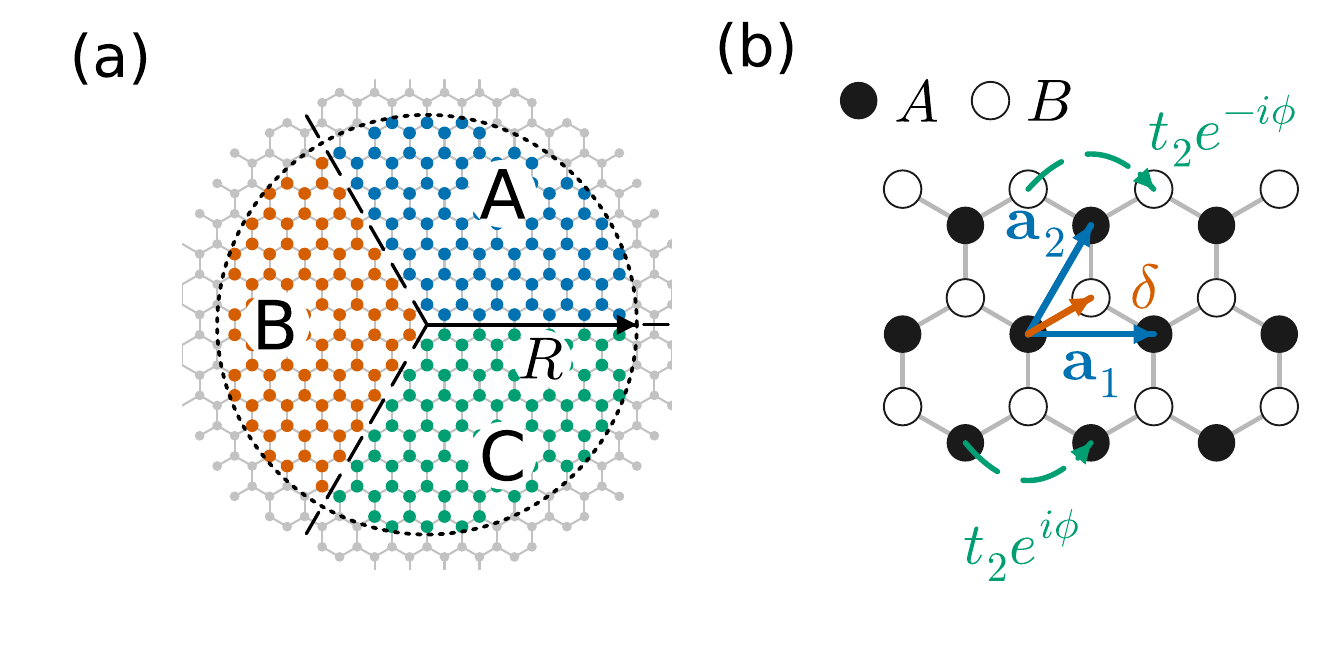}
    \caption{Setup. (a) Tripartition used in this work: three $120^\circ$
    regions $A,B,C$ of a disk with radius $R$, meeting at a plaquette center.
    (b) Conventions on the Haldane honeycomb model: the two sublattices, the Bravais vectors $\bm a_1=(1,0)$ and $\bm a_2=(1,\sqrt{3})/2$, nearest-neighbor vectors $\bm\delta_j$, the nearest-neighbor hopping $t$ (not labeled), and the next-nearest-neighbor hopping $t_2e^{\pm i\phi}$. }
    \label{fig:setup}
\end{figure}

\paragraph*{Modular commutator.}
Given a state $\ket{\psi}$ of a system, and a tripartition $A,B,C$ of a subregion, the MC is defined as:
\begin{equation}
  \JJ \;\equiv\; i\,\langle\psi|\,[K_{AB},K_{BC}]\,|\psi\rangle .
  \label{eq:Jdef}
\end{equation}
where the modular Hamiltonian for a subregion $X$ is given by $K_X\equiv-\log\rho_X$. The MC is real and odd under time-reversal. For a gapped ground
state, and for a tripartition of a disk whose three regions meet at an interior point (Fig.~\ref{fig:setup}(a)),
the MC has been shown to be related to the CCC~\cite{KSKA2022,KSKA2022long}
\begin{equation}
  \JJ \;=\; \frac{\pi}{3}\,\cm .
  \label{eq:KSKA}
\end{equation}
The derivation is an entanglement bootstrap argument~\cite{ShiKatoKim2020}, only assuming entanglement area law of the gapped ground state. Our goal in this work is to generalize the MC to systems with gapless bulk, where the entanglement area law may or may not hold, depending on the exact type of gaplessness.

\paragraph*{Gaussian states.}
We work with free fermions, with the two-point
function given by $\mathcal{G}_{ij}=\langle c_i^\dagger c_j\rangle$. Since there are no higher-order correlations, the two-point correlation matrix $\mathcal{G}$ carries complete information about
the state. Restricting $\mathcal{G}$ to a region $X$ gives $\mathcal{G}_X$, and the
reduced state is Gaussian with single-particle modular
Hamiltonian given by~\cite{Peschel2003,PeschelEisler2009}
\begin{equation}
  K_X \;=\; \log\!\big[(1-\mathcal{G}_X)/\mathcal{G}_X\big] ,
  \label{eq:Kmat}
\end{equation}
with which the MC in Eq.~(\ref{eq:KSKA}) can be readily calculated.
Unless stated otherwise we work on an $L\times L$ torus, so that no physical edge is
present and $\JJ$ is a genuine bulk quantity; the cylinder geometry is used only where
edge modes are explicitly at issue.

\paragraph*{Model.}
We use Haldane's honeycomb model~\cite{Haldane1988}: nearest-neighbor (NN) hopping $t$, next-nearest-neighbor (NNN) hopping $t_2e^{\pm i\phi}$ with opposite phase on the two sublattices, and a staggered onsite potential $\pm M$ (Fig.~\ref{fig:setup}(b)). Writing $\kv$ for crystal momentum, $\sv$ for the Pauli matrices in sublattice space, the Hamiltonian is $h(\kv)=\dv(\kv)\cdot\sv$ with
  $d_x+id_y=-t\!\sum_{j=1}^{3}\!e^{-i\kv\cdot\bm\delta_j}$ and $d_z=M+2t_2\sin\phi\!\sum_{j=1}^{3}\!\sin(\kv\cdot\bm a_j)$, where $\bm\delta_j$ and $\bm a_j$ are the NN and NNN vectors. We fix $t=1$, $\phi=-\pi/2$, $t_2=0.2$ unless otherwise stated. The gapless points of graphene sit at the inequivalent Brillouin-zone corners $K$ and $K'$, where $d_x=d_y=0$. $h(K/K')=(M \mp 3\sqrt{3}\,t_2\sin\phi)\sigma^z$, where the upper (lower) sign referring to $K$ ($K'$). The model realizes $\ch=0,\pm1$, with transitions at $M=\pm M_c$, where $M_c \equiv 3\sqrt{3}\,t_2\sin\phi$. 

We will be looking at the case with a single bulk Dirac node by tuning the original Haldane model to criticality, the case with coexisting chiral edge mode and a bulk Dirac node by considering a double-layer honeycomb, the case with quadratic node using modified hopping, and the case with Fermi surface by doping. 

\begin{figure}[!tbp]
    \centering
    \includegraphics[width=\linewidth]{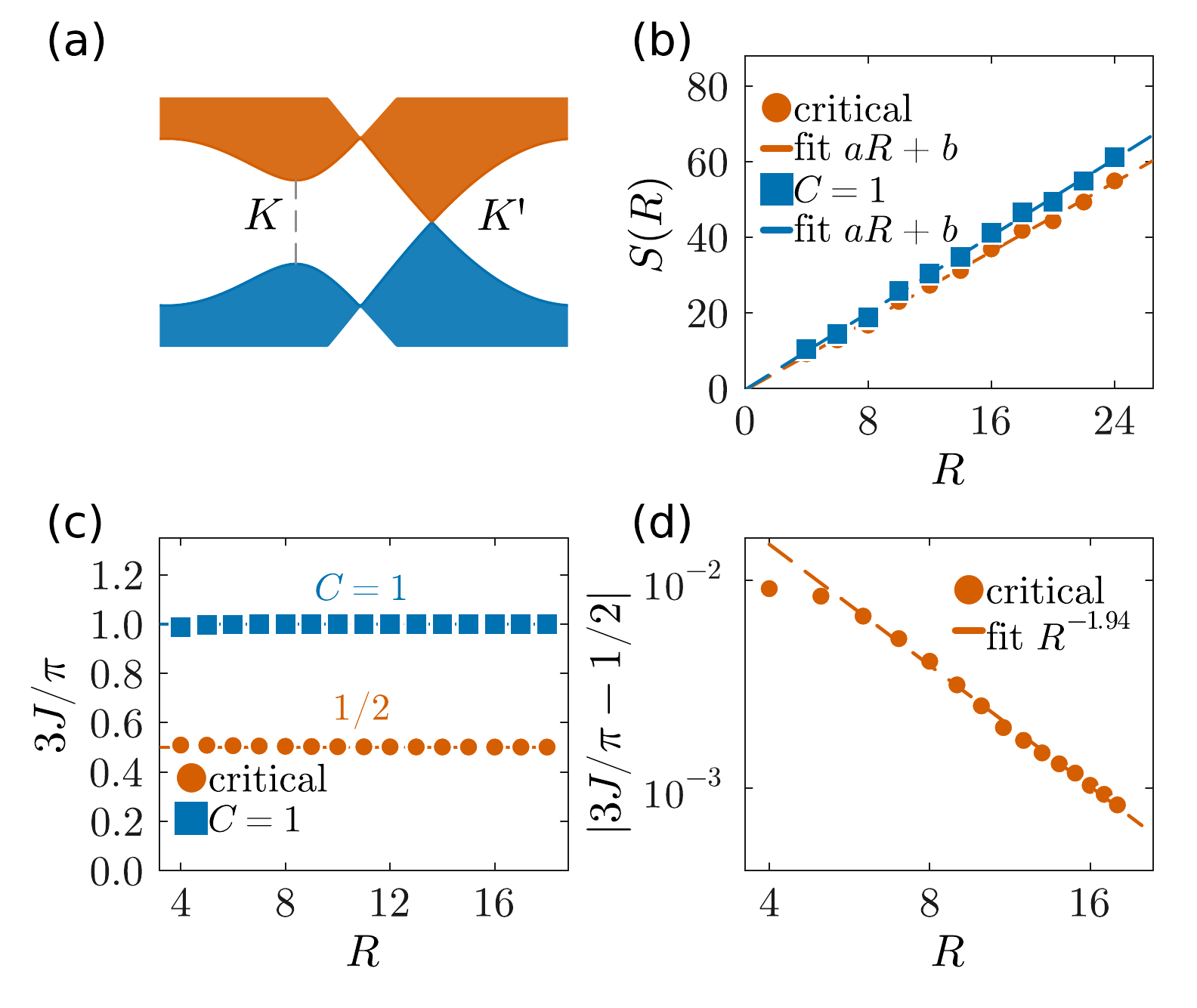}
    \caption{Transition between the $\ch=0$ and $\ch=1$ phases. $M=-0.6$ is used for the gapped case.
    (a) Schematic spectrum at the transition, with one gapless node and one gapped node.
    (b) Entanglement entropy of a disk of radius $R$. Both states follow the area law.
    (c) MC for the tripartition of Fig.~\ref{fig:setup}(a). The
    gapped state returns the Chern number, whereas the critical point converges to 0.5.
    (d) Correction of MC decays as power law in system size.}
    \label{fig:dirac}
\end{figure}
\section{Single Dirac node}
\subsection{Entanglement area law}
The original derivation of Eq.~\eqref{eq:KSKA}  in the gapped case assumes the entanglement area law. Here we check the entanglement scaling numerically.
Fig.~\ref{fig:dirac}(b) shows the entanglement entropy of a disk of radius $R$ is linear in $R$, for both the gapless case (transition between the Chern number $\ch=1$ and $\ch=0$ phases) and the gapped case ($\ch=1$). This is expected. For free fermions the Widom formula predicts a multiplicative logarithmic violation only when the Fermi sea has an extended boundary~\cite{calabrese2004entanglement,Gioev2006,Wolf2006,Swingle2010}, whereas an isolated point node in two dimensions the area law remains intact~\cite{fradkin2006entanglement}, with corrections that are subleading and power-law in $R$. The gapless bulk with a single Dirac node therefore does not spoil the area law requirement, even though it is a different kind of area law from the one in the original entanglement bootstrap program for the gapped phases~\cite{ShiKatoKim2020,KSKA2022,KSKA2022long,li2025strict-area}. This is indicated by the violation of the Markov property~\cite{KSKA2022} shown in Appendix~\ref{append-sec:markov}.

\subsection{Half-quantized MC and parity anomaly}
Fig.~\ref{fig:dirac}(c) plots the measured MC in different cases. The non-trivial gapped case simply gives the Chern number, which in this case is just the CCC $c_-$, agreeing with Eq.~(\ref{eq:KSKA}). Quite interestingly, the MC (scaled by $3/\pi$) converges to $1/2$ at the critical point with a single Dirac cone, even though there is no chiral (Majorana) edge mode. Fig.~\ref{fig:dirac}(d) further shows the deviation of the MC from its limiting value $1/2$ decays as a power law with subsystem size, which is reasonable given the power law correlations in gapless systems. We show in Appendix~\ref{append-sec:deform} that the half-quantization is robust under tripartition deformation, as well as when the Dirac velocity and the Dirac cone anisotropy is tuned.

Next we provide a heuristic symmetry argument to show that the MC of the system with a single Dirac cone is half-integer quantized. More precisely, given the Chern numbers $\ch_\pm$ of the two nearby gapped phases, with $\Delta\ch\equiv \ch_+-\ch_-=1$ such that only one gapless Dirac cone at the transition, we claim (in thermodynamic limit):
\begin{equation}
    \label{eq:claim}
    3J/\pi=\frac{\ch_++\ch_-}{2}=\ch_-+\frac{1}{2}\in\mathbb{Z}+\frac{1}{2}.
\end{equation}
\begin{proof}
We assume the generic situation where the transition is achieved by tuning an effective mass parameter $m$ of the low energy Dirac fermion, with $\ch(m=\pm\delta)=\ch_\pm$, where $\delta$ is a small positive mass, and the transition happens at $m=0$, where $\ch$ becomes ill-defined. Now we make a physically motivated assumption: 
\begin{equation}
    3J(m)/\pi=\Phi_h(m)+\Phi_l(m),
    \label{eq:scale-splitting}
\end{equation}
where $\Phi_{h/l}$ represents the contributions to the MC from the high/low-energy modes (see Appendix~\ref{append-sec:proof} for more details). $\Phi_l$ comes from the Dirac cone part that's approximately described by the Hamiltonian $h(m)\simeq v_xq_x\sigma^x+v_yq_y\sigma^y+m\sigma^z$ (since the argument only involves reflection symmetry, anisotropy is also allowed). We argue that the low-energy contribution $\Phi_l(m)$ is odd in the mass $m$. Firstly, notice that $m\to-m$ is implemented in the infrared by a spatial reflection (say $q_y\to-q_y$), realized by the parity transformation on the Dirac field $\psi(q_x,q_y)\to\sigma^x\psi(q_x,-q_y)$. 
Under parity, $h(m)\to h(-m)$, with the corresponding contribution to the MC given by $\Phi_l(-m)=-\Phi_l(m)$, where the negative sign in front of $\Phi_l$ comes from the reversed orientation of the tripartition, under which the MC flips sign. Now using Eq.~(\ref{eq:scale-splitting}), we have
\begin{equation}
\label{eq:sum}
    3/\pi(J(\delta)+J(-\delta))=\Phi_h(\delta)+\Phi_h(-\delta)\xrightarrow{\delta \to 0^+}2\Phi_h(0),
\end{equation}
where $J(\pm\delta)$ are readily related to $\ch_\pm$ of the two nearby gapped phases, and the assumption that the high-energy contribution to the MC is continuous across the transition is made (see Appendix~\ref{append-sec:proof}). The claim in Eq.~(\ref{eq:claim}) follows immediately. Similarly, by subtraction instead of summation as in Eq.~(\ref{eq:sum}), we obtain $\Phi_l(0^\pm)=\pm1/2=\frac{1}{2}\mathrm{sign}(m)$, even though we also have $\Phi_l(0)=0$ from the oddness of the function. It is apparent that $\Phi_l(m)$ is not continuous near $m=0$ in the thermodynamic limit, and the discontinuity is the essence of the Chern number changing transition.
\end{proof}
The same half-integer governs the transport response of a
single massive Dirac fermion on the 2d surface of a 3d topological insulator~\cite{FuKane2007,senthil-review2015}:
the Hall conductance $\sigma_{xy}=\tfrac{e^2}{2h}\,\mathrm{sign}(m)$. This is also related to the 2d parity 
anomaly of Dirac fermions~\cite{Niemi1983,Redlich1984,Semenoff1984,jackiw1986}: no regularization preserves both gauge invariance and parity. The partition function of a 2d massless Dirac fermion coupled to a background $U(1)$ gauge field acquires the phase $e^{-i\pi\eta/2}$, with $\eta$ the Atiyah--Patodi--Singer invariant of the Dirac operator~\cite{AGDPM1985,Witten2016,APS1975}, which essentially keeps track of the number of eigenvalues of the Dirac operator that change sign during a gauge transformation. A jump of 2 in $\eta$ would flip the sign of the partition function, leading to gauge non-invariance. On a lattice the anomaly is resolved as in Haldane's
construction~\cite{Haldane1988}. At the transition between the topologically trivial ($\ch=0$) and non-trivial ($\ch=\pm1$) phases, one Dirac node remains gapless whereas the other becomes massive, whose mass term breaks time-reversal and parity. This massive partner acts
as the physical Pauli--Villars regulator for the low-energy gapless Dirac fermion, so that the system as a whole is anomaly-free. 

In this language, the half-quantized MC probes exactly the parity-breaking effect of the massive Pauli--Villars partner of the massless Dirac fermion. The corresponding half-quantized Hall conductance has recently been observed in semi-magnetic topological-insulator films~\cite{Mogi2022}. What our main result in Eq.~(\ref{eq:claim}) adds to the understanding of the parity anomaly is that the same half-quantization is also visible in an \textit{entanglement
property} of the ground-state wavefunction. With regard to the Hall conductance, we note in passing that a recently proposed formula~\cite{FanSahayVishwanath2023} to extract the Hall conductance using a single wavefunction also works for the gapless Dirac node case and gives the right half-quantized value (see Appendix~\ref{append-sec:hall} for details).
\section{Dirac node coexisting with chiral edge mode}
\begin{figure}
    \centering
    \includegraphics[width=\linewidth]{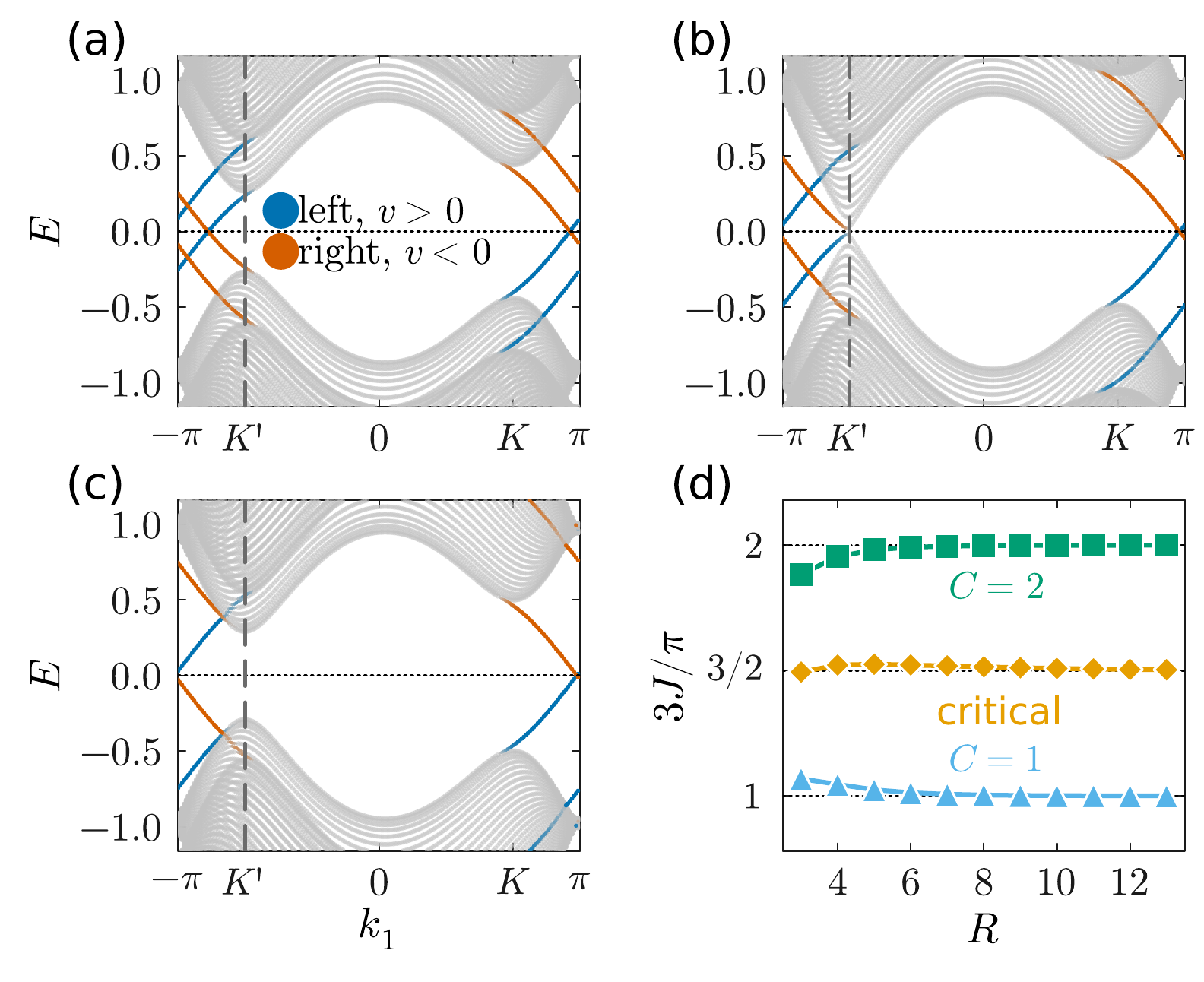}
    \caption{Chiral edge mode coexisting with a bulk Dirac node, Eq.~\eqref{eq:bilayer}, at
    $t_c=0.3$ and $M_1=0$.
    (a)-(c) Cylinder spectra (periodic along $\bm a_1$). Grey marks
    bulk states; colored points are edge modes (the higher the color intensity, the more localized to the edge) , blue for left-moving ($v>0$) and
    orange for right-moving ($v<0$). The dashed line is the projection of $K'$.
    (a) $\ch=2$: two pairs of co-propagating chiral edge modes.
    (b) At $M_2=M_2^c$ the bulk closes, where one pair of edge modes delocalize into the bulk and the other pair remains robust.
    (c) $\ch=1$: one pair of edge modes.
    (d) $3J/\pi$ versus $R$ for the three cases in (a)-(c): converges to $2$, $3/2$, and $1$, respectively.}
    \label{fig:coexist}
\end{figure}
The case with a single Dirac cone discussed above has no chiral edge mode. To relate back to the connection between the MC and the CCC $c_-$ in the gapped case~\cite{KSKA2022,KSKA2022long}, we consider the situation where a robust chiral edge mode coexists with the gapless bulk (e.g. an example of gapless SPT). A simple free-fermion realization can be achieved by coupling two layers of the Haldane honeycomb model (several similar constructions exist in the literature~\cite{Verresen2018,Verresen2020,GuoYangYu2026,zeng2026}). The idea here is to realize a $\ch:2\to1$ transition, such that the critical point has one Dirac cone in the bulk, as well as a chiral edge mode when there is an open boundary. More precisely, we consider the Hamiltonian:
\begin{equation}
  H(\kv)=\begin{pmatrix} h(\kv;M_1) & t_c\,\mathbbm 1\\ t_c\,\mathbbm 1 & h(\kv;M_2)\end{pmatrix},
  \label{eq:bilayer}
\end{equation}
with $M_1=0$, so that layer one sits at $\ch=1$, and $M_2$ tunable.
The gap therefore closes at $M_2^{c} \;=\; M_c-\frac{t_c^{2}}{M_c}$. The partner valley $K$ stays gapped, so the low-energy bulk at $M_2=M_2^c$ is again a single massless Dirac cone, where one pair of chiral/anti-chiral modes merge into the bulk, and the other pair survive. Fig.~\ref{fig:coexist}(a)-(c) confirms this on the cylinder. The surviving chiral edge branches cross zero energy at a momentum well away from the bulk nodal point. Edge and bulk gaplessness thus coexist at different conserved momenta, which is what makes the state a free-fermion analog of a gapless SPT. Fig.~\ref{fig:coexist}(d) shows that $3J/\pi$ converges to the respective Chern numbers in the two gapped phases, as expected, and  at the critical point it converges to the average of the two, agreeing with the prediction in Eq.~(\ref{eq:claim}). The MC now has contribution from both the critical bulk and the chiral edge, measuring the total chirality of the system.

\section{Other types of gaplessness}
\begin{figure}
    \centering
    \includegraphics[width=\linewidth]{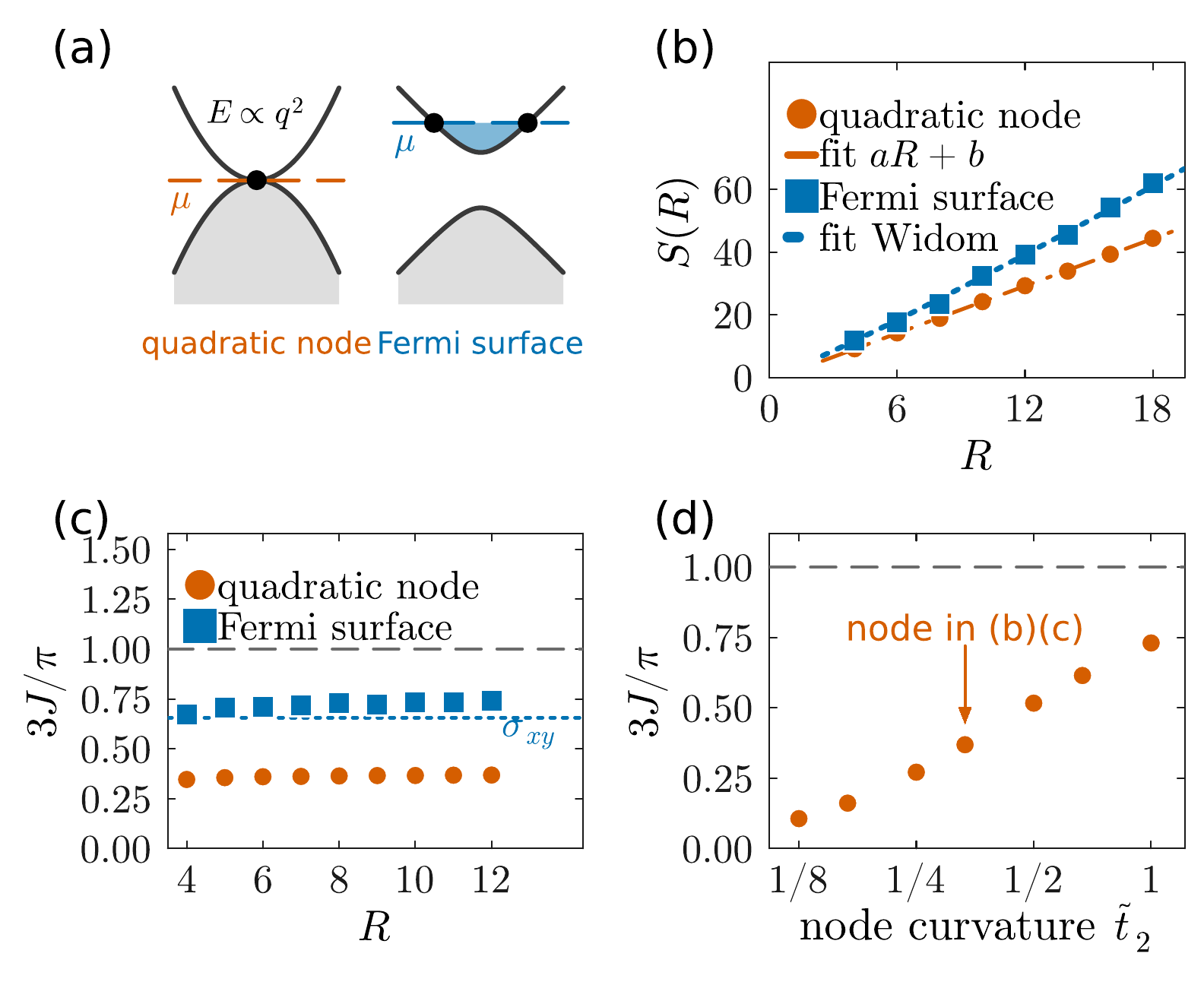}
    \caption{(a) Schematic figures for quadratic node (left) and finite Fermi surface with $\mu=0.75$ (right). (b) Entanglement scaling: Quadratic node still satisfies the area law, whereas the FS follows the Widom scaling~\cite{Gioev2006}. (c) The MC: both cases seem to converge in the thermodynamic limit, but we note that the quadratic node case differs significantly from $1=(\ch_++\ch_-)/2$, and the FS case differs from the Berry curvature $\sigma_{xy}$ (dotted line)~\cite{haldane2004berry}. (d) Curvature dependence of the MC for a quadratic node.}
    \label{fig:other-gapless}
\end{figure}
Here we further investigate other types of gaplessness, using the quadratic node and the Fermi surface as examples. Firstly, to achieve a quadratic node, we use the following variant of the Haldane model:
\begin{equation}
  h_n(\mathbf k)=\begin{pmatrix} d_z(\mathbf k) & w_n(\mathbf k)\\[2pt]
                                 w_n^*(\mathbf k) & -d_z(\mathbf k)\end{pmatrix},
  \quad
  w_n(\mathbf k)=-\tilde{t}_n\,f^n(\mathbf k),
  \label{eq:hn}
\end{equation}
with $f(\bm k)=\sum_{j=1}^{3}\!e^{i\kv\cdot\bm\delta_j}$. For $n=1$, Eq.~\eqref{eq:hn} is the Haldane model. Taking $n=2$ doubles the winding, realizing a $\ch:2\to0$ transition with a quadratic band touching $E\propto q^2$ at the critical point.  Here we take $\tilde{t}_1=t=1$ and $\tilde{t}_2=1/3$. Secondly, to achieve a Fermi surface we use a non-zero chemical potential $\mu$ on top of the original gapped Haldane model (here only $\tilde{t}_1$ term is kept). The schematics for the two cases are shown in Fig.~\ref{fig:other-gapless}(a). 

Fig.~\ref{fig:other-gapless}(b) shows the entanglement scalings. The quadratic node is fitted with a linear function, and the Fermi surface is fitted with the Widom form $S(R)\sim R\ln R+\mathcal{O}(R)$~\cite{Gioev2006}, though we have to note that at relatively small system sizes, the difference between the Widom and the linear is difficult to distinguish. Fig.~\ref{fig:other-gapless}(c) shows the measured MCs for both cases, which are no longer (half-)quantized. In fact, we show in Appendix~\ref{append-sec:deform} that the MC for Fermi surface is no longer invariant under the deformation of the tripartition. In that sense, the MC is not well-defined for Fermi surfaces (even though we still labeled the Fermi surface Hall conductance $\sigma_{xy}$ as a reference~\cite{haldane2004berry}). In contrast, the MC of the quadratic node does not depend strongly on the geometries of the tripartition being tested. However, we do note that the curvature of the quadratic node does play a role (Fig.~\ref{fig:other-gapless}(d)).

\section{Conclusion and outlook}
We have investigated the MC for bulk gapless systems with different types of gaplessness, going beyond the original gapped formulation. Most interestingly, in the presence of a massless Dirac cone the MC is half-quantized, the entanglement manifestation of the parity anomaly. In this respect, the Dirac cone is a special kind of gaplessness due to its robust universal low energy physics.

Several directions remain open. All results reported here are for free fermions, where the correlation matrix gives direct access to modular Hamiltonians. The symmetry argument behind the half-quantization does not, however, rely on Gaussianity, and we expect it to
hold even when interactions are included; testing this in an interacting gapless state would be a natural next step, though the numerical cost in such settings is expected to be significant. However, there are recent progress in Monte Carlo algorithm on extracting the modular Hamiltonian~\cite{yan2025MC-RDM}. Furthermore, we regard a first-principle 
demonstration of the relation between the half-quantized MC and the $\eta$-invariant of the Dirac operator as another important open problem. Establishing this would require a derivation directly from the modular Hamiltonians of the tripartition, which we leave for future study.

\emph{Note added.} As this paper is being completed, we noticed a recent paper by Moy and Fradkin~\cite{moy2026chern} that studied similar physics from a different perspective by explicitly looking at edge correlation functions and anomaly inflow. The two works are therefore complementary to each other.
\section*{Acknowledgments}
MZ would like to thank Bowen Shi and Lei Su for insightful discussions. This work is supported in part by the Alexander von Humboldt Foundation through a Humboldt Research Fellowship. 

\bibliography{references.bib} 
\clearpage
\appendix
\section{Violation of Markov property}
\label{append-sec:markov}
\begin{figure}
    \centering
    \includegraphics[width=\linewidth]{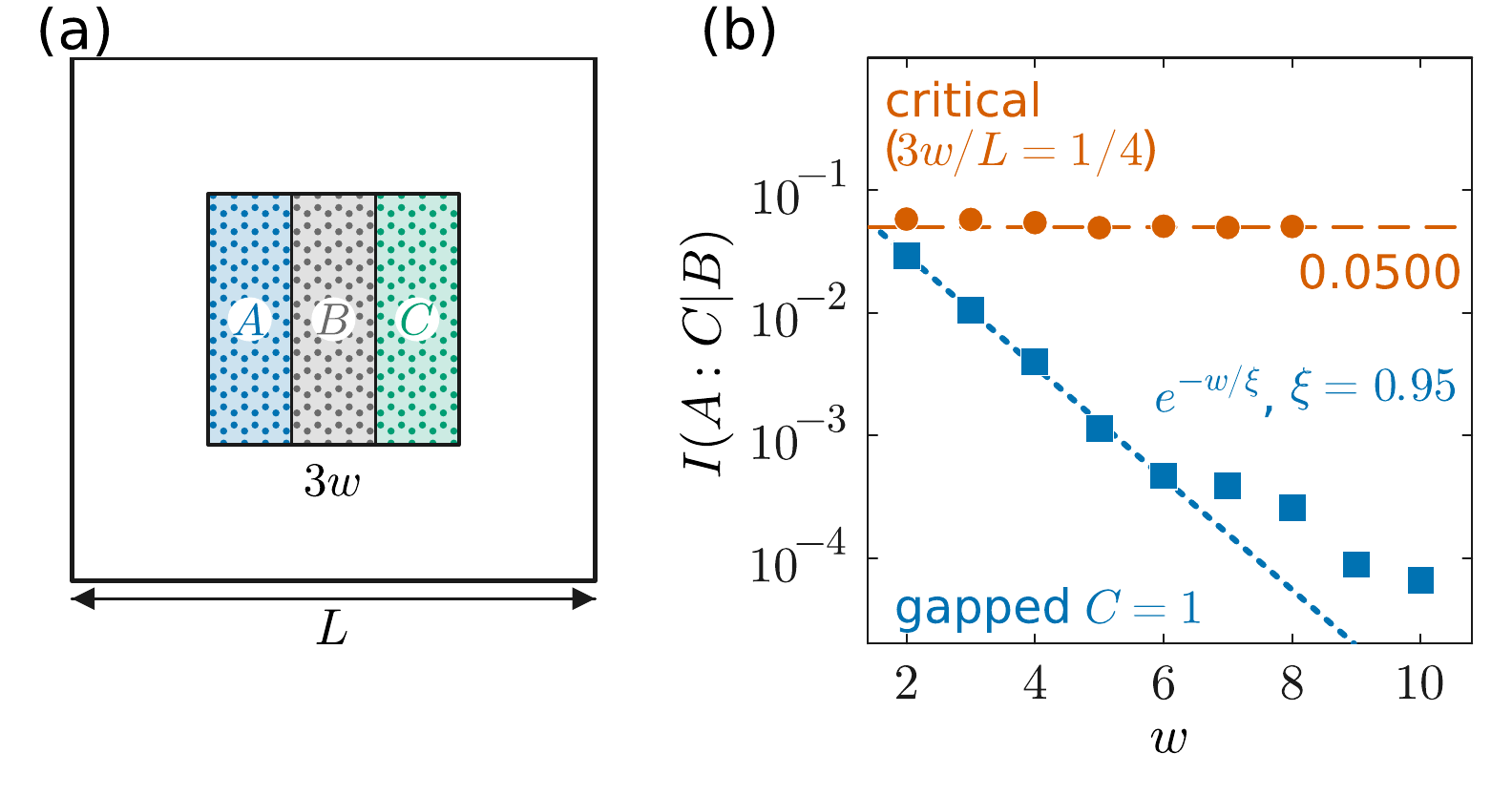}
    \caption{Markov property check. (a) The chain tripartition $ABC$ of a $3w\times 3w$ square. (b) CMI: decays exponentially in the gapped case and saturates in the critical case (with a Dirac node).}
    \label{fig:cmi}
\end{figure}
The Markov property is defined through the conditional mutual information (CMI) of a chain partition of a subregion of a state (see Fig.~\ref{fig:cmi}(a), i.e. $A$ and $C$ are fully separated from one another by $B$). Writing the von Neumann entanglement entropy $S_X=-\Tr\rho_X\log\rho_X$, the CMI is defined by
\begin{equation}
  I(A\!:\!C|B) \;=\; S_{AB}+S_{BC}-S_{B}-S_{ABC}.
  \label{eq:cmi}
\end{equation}
The Markov property states that the CMI for a gapped state must decay to zero exponentially with the size $w$. Operationally the Markov property is the statement that the modular Hamiltonian is local, in the sense that it decomposes additively, $ K_{ABC}\simeq K_{AB}+K_{BC}-K_B$~\cite{HJPW2004}.

Fig.~\ref{fig:cmi}(b) shows that the CMI of the critical point saturates when the (sub)system sizes are proportionally increased, in stark contrast to the exponential decay of the gapped case. This implies that even though the entanglement area law is still satisfied to leading order at the critical point with a Dirac node, it is a different type of area law assumed in the original entanglement bootstrap program for the gapped phases, because the latter already implies the Markov property~\cite{ShiKatoKim2020,li2025strict-area}.

\section{Robustness of the MC: geometry, Dirac velocity and anisotropy}
\label{append-sec:deform}
\begin{figure}
    \centering
    \includegraphics[width=\linewidth]{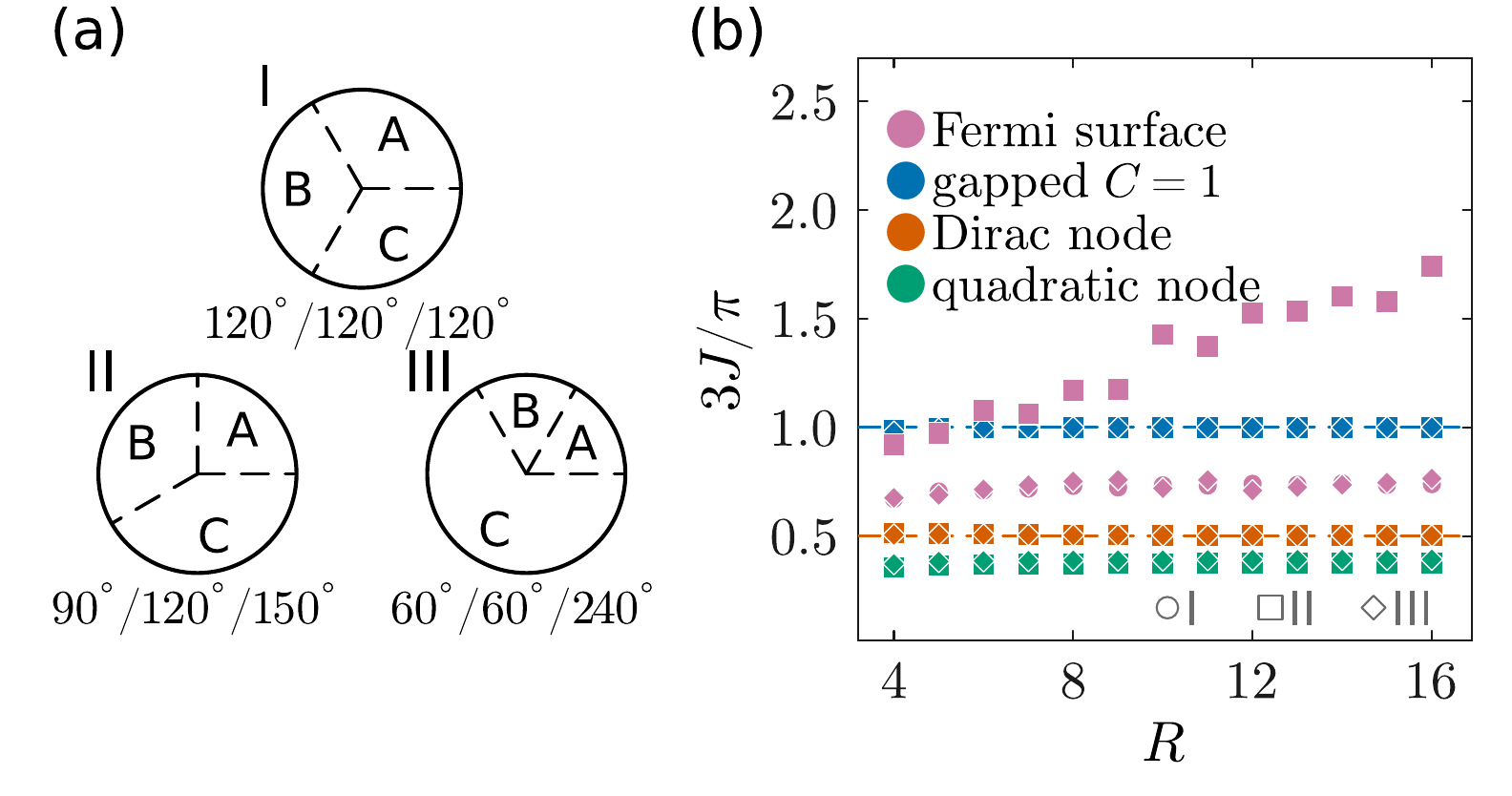}
    \caption{Deformation invariance check. (a) Different geometries of tripartition. (b) The (non)-convergence of the MC in various cases under tripartitions given in (a).}
    \label{fig:deform}
\end{figure}
\begin{figure}
    \centering
    \includegraphics[width=\linewidth]{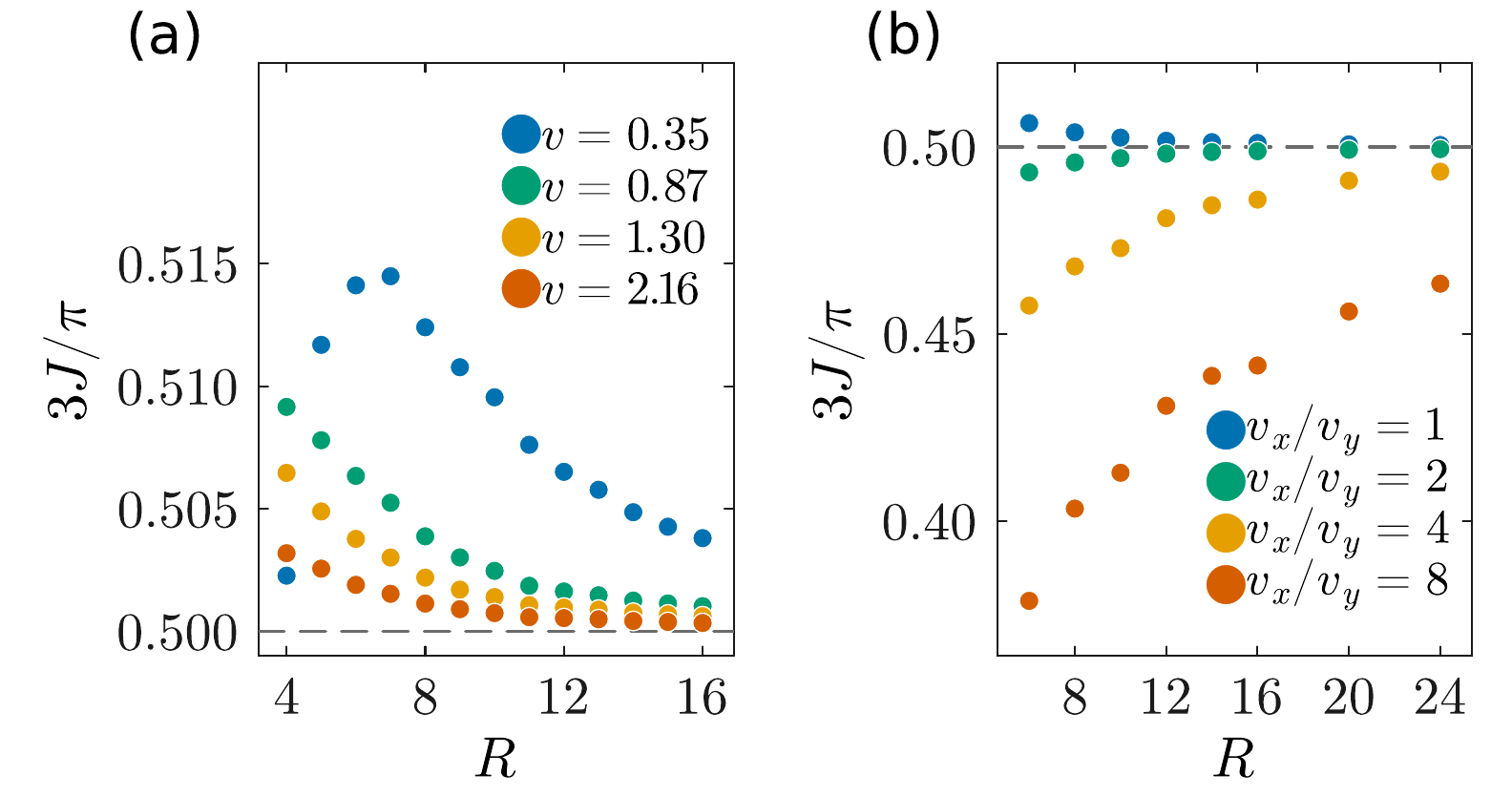}
    \caption{Effects of linear velocity (a) and anisotropy (b) on the MC of states with a Dirac cone.}
    \label{fig:robust}
\end{figure}
In this section, we check the robustness of the MC in various cases. Fig.~\ref{fig:deform}(a) display the three different geometries of the tripartition, and Fig.~\ref{fig:deform}(b) shows the measured MC. The case with Fermi surface shows strong dependence on the geometry, whereas the gapped case and the two cases with point nodes remain robust for the tested geometries. We suspect that the behavior of the Fermi surface is directly related to the violation of the entanglement area law.

Furthermore, we check the robustness of the MC for the Dirac cone case when the linear velocity is tuned, and when the cone becomes anisotropic. Fig.~\ref{fig:robust} shows that the half-quantization remains robust in both cases.
\section{Fan-Sahay-Vishwanath formula for Hall conductance}
\label{append-sec:hall}
\begin{figure}
    \centering
    \includegraphics[width=\linewidth]{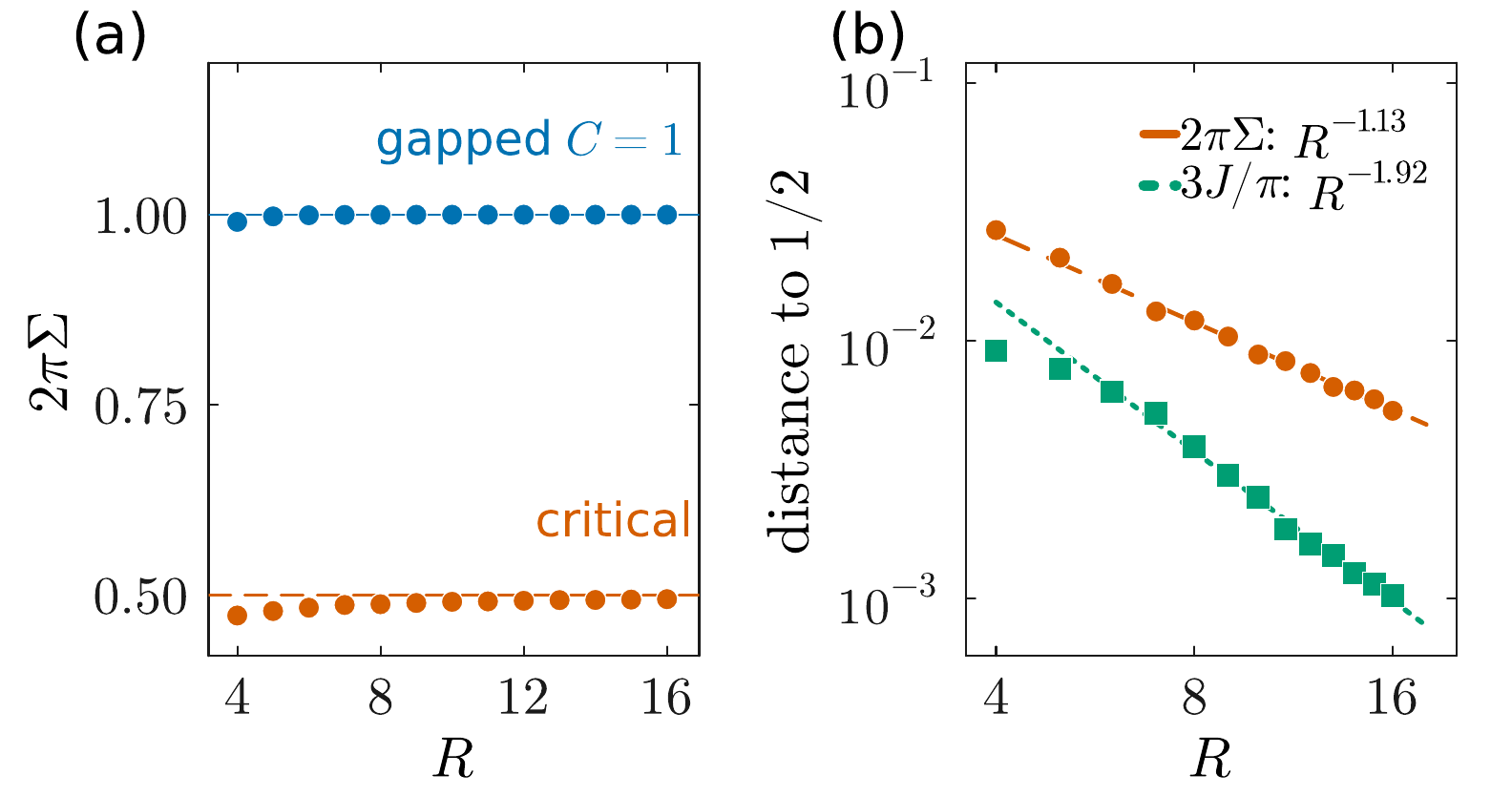}
    \caption{(a) The FSV formula applied to the gapped and the critial cases. (b) The convergence of the FSV formula in the critical case, with the MC also shown for comparison. Both follow a power law.}
    \label{fig:hall}
\end{figure}
Recently, Fan, Sahay and Vishwanath proposed a new formula (referred to as the FSV formula below) to extract the Hall conductance of a gapped state using a single bulk wavefunction~\cite{FanSahayVishwanath2023}. Given the same tripartition used for the modular commutator, the FSV formula says
\begin{equation}
    \Sigma(A,B,C)\equiv \frac{i}{2}\langle[K_{AB},Q_{BC}^2]\rangle=\sigma_{xy},
    \label{eq:fsv}
\end{equation}
where $K_{X}$ is defined as before, and $Q_X$ is the total charge operator in the region $X$, $\sum_{j\in X}c_j^\dagger c_j$. 

Here we apply the FSV formula to the critical point with a Dirac cone. Fig.~\ref{fig:hall} shows that, similar to the MC, it is also half-quantized, and both quantities have power law convergence with system size.

\section{Energy scale separation of the MC}
\label{append-sec:proof}
Given the Hamiltonian with the mass parameter $m$, we denote the many-body eigenstates as ${\ket{\phi_n}}$, with $n=0$ labeling the ground state (GS), which is assumed to be non-degenerate. Here we make the dependence on $m$ implicit. The GS density operator of the whole system is given by $\rho=\ket{\phi_0}\bra{\phi_0}$. Now consider the small variation $m\to m+\dm$, and denoting the corresponding mass operator by $\hat{O}$. Then the perturbed ground state is given by
\begin{equation}
    \ket{\tilde{\phi}_0}=\ket{\phi_0}+\dm\sum_{n\neq 0}O_{0n}\ket{\phi_n}+\mathcal{O}(\dm^2),
\end{equation}
where $O_{0n}\equiv\frac{\bra{\phi_n}\hat{O}\ket{\phi_0}}{E_0-E_n}$. Then the new density operator is given by
\begin{equation}
\label{eq:drho}
    \tilde{\rho}=\rho+\dm \sum_{n\neq 0}\left(O_{0n}\ket{\phi_n}\bra{\phi_0}+h.c.\right)+\mathcal{O}(\dm^2).
\end{equation}
From now on we only keep the leading linear terms in $\dm$. Then for any reduced density matrix on region $X$, we have
\begin{equation}
\label{eq:variation}
\begin{split}
    &\delta \rho_X=\delta\Tr_{\bar{X}}\rho=\Tr_{\bar{X}}\delta \rho,\\
    &\delta K_X=-\delta\log\rho_X=-\int_0^\infty dt(\rho_X+t)^{-1}\delta \rho_X(\rho_x+t)^{-1},
    \end{split}
\end{equation}
where the variation in the modular Hamiltonian $K_X$ is given by the Fréchet derivative. The point is that, all the variations are linear in $\delta \rho$.
Since
\begin{equation}
\begin{split}
    \delta J=&i\Tr\left(\delta \rho [K_{AB},K_{BC}]\right)+i\Tr\left( \rho [\delta K_{AB},K_{BC}]\right)\\
    &+i\Tr\left(\rho [K_{AB},\delta  K_{BC}]\right),
    \end{split}
\end{equation}
where the variations of different terms can be related to $\delta \rho$ based on Eq.~(\ref{eq:variation}), we have
    $\delta J=i\Tr\left(\delta\rho \mathcal{M}\right)$,
where $\mathcal{M}$ represents a constant matrix with complicated dependence on the various reduced density matrices and modular Hamiltonians, whose exact form is not important for our discussion. With $\delta \rho$ given in Eq.~(\ref{eq:drho}), we eventually arrive at
\begin{equation}
    \delta J=\dm \sum_{n\neq 0}\left(\frac{\bra{\phi_n}\hat{O}\ket{\phi_0}\bra{\phi_0}\mathcal{M}\ket{\phi_n}}{E_0-E_n}+h.c.\right).
\end{equation}
As advertised, we can choose a cutoff $\Lambda$ in energy to divide the contributions to high-energy and low-energy parts. Writing $\varepsilon_n\equiv E_n-E_0$ for the excitation energies, the cutoff splits the response as $\mathcal{Q}(m)\equiv\frac{\partial J}{\partial m}=\Ql(m)+\Qg(m)$, with
\begin{equation}
\label{eq:Gsplit}
\mathcal{Q}_{\lessgtr}=\!\!\!\sum_{n\neq0,\;\varepsilon_n\lessgtr\Lambda}\!\!\!
\left(\frac{\bra{\phi_n}\hat{O}\ket{\phi_0}\bra{\phi_0}\mathcal{M}\ket{\phi_n}}{E_0-E_n}+h.c.\right),
\end{equation}
with $\Lambda$ in the window $1/R\ll\Lambda\ll\Delta_{\rm UV}$, where $R$ is the linear size of the tripartition and $\Delta_{\rm UV}$ the lattice scale. Every denominator in $\Qg$ obeys $\varepsilon_n\ge\Lambda$, so the high-energy branch is manifestly free of the singularity that produces the transition.

Up to now, everything is general and exact. To get more physical intuition, we limit ourselves to free fermion systems, where $J$ becomes a functional of the two-point correlation function, such that $\mathcal{Q}(m)=\partial_m J=\frac{\delta J}{\delta \mathcal{G}}\partial_m \mathcal{G}$. Heuristically, the form of $\mathcal{G}$ depends on the energy scale. We focus on the neighborhood of the transition, and can take $|m|\ll 1/R$. At scale below $\Lambda$, we have the Dirac form
\begin{equation}
    \mathcal{G}(r)\sim e^{-mr}/r^2.
\end{equation}
In contrast, for scales above $\Lambda$, there is an effective gap of the order $\Lambda$ for the excitations, and the correlation can be approximated by $\mathcal{G}(r)\sim e^{-\Lambda r}$, which becomes manifestly $m$-independent and leads to $\mathcal{Q}_>(m)\to 0$. Since $\mathcal{Q}$ is the derivative of $J$, the high-energy background is effectively constant near the transition. Now for the low-energy part, integrating across the regions with linear size $R$, we have 
\begin{equation}
    \mathcal{Q}_<(m)\sim \int_0^Rdr r \partial_m \left(\frac{e^{-mr}}{r^2}\right)\sim \frac{1-e^{-mR}}{m}\xrightarrow{m \to 0} R,
\end{equation}
which diverges in the thermodynamic limit $R\to \infty$. Namely, the discontinuity of $J$ comes from the low-energy part. When $R$ is finite, we have 
\begin{equation}
    J(m)\sim \int_0^m dm'\mathcal{Q}_<(m')+\int_0^m dm'\mathcal{Q}_>(m')+J(0),
\end{equation}
where the 1st term can be identified with $\Phi_l(m)$ and the last two terms with $\Phi_h(m)$ in the main text.
\end{document}